\documentclass[twocolumn,showpacs,preprintnumbers,amsmath,amssymb]{revtex4}

\usepackage{graphicx}
\usepackage{dcolumn}
\usepackage{bm}
\input epsf

\begin{document}

\preprint{}

\title{Localization of interacting fermions at high temperature}
\author{Vadim Oganesyan} \email{vadim.oganesyan@yale.edu}
\affiliation{Department of Physics, Yale University, New Haven CT
06520}
\author{David A. Huse}
\email{huse@princeton.edu} \affiliation{Department of Physics,
Princeton University, Princeton NJ 08544}
\date{\today}

\begin{abstract}
We suggest that if a localized phase at nonzero temperature $T>0$
exists for strongly disordered and weakly interacting electrons, as
recently argued, it will also occur when both disorder and
interactions are strong and $T$ is very high.  We show that in this
high-$T$ regime the localization transition may be studied
numerically through exact diagonalization of small systems.  We
obtain spectra for one-dimensional lattice models of interacting
spinless fermions in a random potential.  As expected, the spectral
statistics of finite-size samples cross over from those of
orthogonal random matrices in the diffusive regime at weak random
potential to Poisson statistics in the localized regime at strong
randomness. However, these data show deviations from simple
one-parameter finite-size scaling: the apparent mobility edge
``drifts'' as the system's size is increased. Based on spectral
statistics alone, we have thus been unable to make a strong
numerical case for the presence of a many-body localized phase at
nonzero $T$.
\end{abstract}

\maketitle

\section{\label{sec:intro}Introduction}
Although Anderson's original paper on localization \cite{pwa58} is
mostly remembered for its ground-breaking results about single
particles in random potentials, one goal of that paper was to learn
about transport properties of highly-excited many-body eigenstates,
e.g. quantum diffusion of nuclear moments. This latter goal was
mostly neglected in subsequent research on localization and
metal-insulator transitions. However, these questions have been
recently brought to our attention by Basko, {\it et al.} \cite{baa},
who present detailed arguments that interacting electrons in static
random potentials can have a true metal-insulator transition at a
nonzero critical temperature. Thus these systems are argued to have
an insulating phase, with strictly zero ohmic conductivity, even at
a nonzero temperature. For some work on these questions published
before Basko, {\it et al.}, see for example \cite{fa80, bs99, ss00,
ngld, mnr, dykman, gmp}.

In practice, few transport measurements are possible without first
equilibrating the sample with its environment in order to establish
a steady state (by removing Joule heat). In metals this coupling to
the environment, provided it is not too strong, does not affect the
conductivity (non-linear transport is another story altogether, see
e.g. \cite{tremblay}). In Anderson insulators, however, the heat
bath plays a far less subtle role: it is what permits transport.
Conduction occurs by variable-range-hopping, which is an inelastic
process requiring a heat bath that can locally supply or absorb the
energy needed to permit hopping of the charge carriers between
localized states that are not precisely degenerate. At the heart of
this extreme sensitivity of the dynamics of a localized insulator to
the coupling with its environment is its inability to
self-equilibrate. It is therefore useful to turn the issue around by
distinguishing conductors from true $T>0$ insulators by whether the
many-particle system itself constitutes a heat bath. For example,
one might ask whether external local probes can deposit limitless
amounts of energy or if they tend to saturate the spectrum.
Similarly, whether or not attached leads themselves can effectively
remove heat from the sample will generally depend on heat
conductivity of the sample itself. Thus we see that whether or not a
quantum system of many interacting degrees of freedom constitutes a
heat bath is not only a very fundamental question, but also one of
some practical relevance.

To the extent that one of the most successful theories of nature,
namely thermodynamics, is founded on the assumption of ergodicity,
we expect true insulators (where this assumption is strongly
violated) to be rare and require fine tuning of some sort. The
noninteracting Anderson insulator is one example, where the
unrealistic condition of no interparticle interactions is crucial.
Remarkably, the authors of Ref. 2 argue that a nonzero temperature
Anderson insulator can be stable against the dephasing effects of
interparticle interactions, making this state a sufficiently
realistic possibility to be taken seriously and looked for in
experiments (provided decoherence from the rest of the universe can
be ignored to a good approximation).

The calculations of Ref. 2 are based on a low energy effective
Hamiltonian whose connection with the parameters of the original
model of interacting electrons in a random potential could not be
established analytically.  Thus, it is interesting and likely
worthwhile to test their results using other methods, and to try to
learn more about the nature of the proposed $T>0$
diffusive-to-insulating phase transition and about the range of
models that may exhibit it. We report here on one such attempt. To
start we observe that application of the quantitative estimates of
the localization transition in Ref. 2 to a lattice model with finite
entropy and energy densities (i.e. finite number of states at each
site) implies that the aforementioned localized phase and,
therefore, the phase transition to the diffusive state can persist
all the way to \emph{infinite} temperature.  This seemingly
innocuous observation has at least two important practical
implications. First, by adapting familiar high temperature expansion
techniques we can more or less rigorously rule out the possibility
that such a transition is accompanied by a thermodynamic signature
both at infinite temperature and by continuity at any finite
temperature \cite{SLS}. Perhaps more interestingly, the very large
(exponential in volume) number of states available to the system at
high temperatures can sometimes create favorable conditions for
quickly approaching the thermodynamic limit in various thermodynamic
\emph{and} dynamic quantities, which raises the possibility of
looking for the signs of this physics numerically, e.g. in exact
spectra of finite samples. Our choice of the model and method of
analysis are summarized below, followed by results and some
preliminary conclusions.

\section{Ensemble of Hamiltonians}
To reach as large as possible a distance with a given size many-body
Hilbert space, we study spinless fermions hopping and interacting on
a one-dimensional lattice of $L$ sites with a random potential and
periodic boundary conditions.  This model has only two states (empty
and occupied) per lattice site. The Hamiltonian is
\begin{eqnarray}
H=\sum_i[w_in_i + V(n_i-{{1}\over{2}})(n_{i+1}-{{1}\over{2}})
\nonumber \\
+
c^{\dagger}_ic_{i+1}+c^{\dagger}_{i+1}c_i+c^{\dagger}_ic_{i+2}+c^{\dagger}_{i+2}c_i]~.
\end{eqnarray}
The nearest-neighbor interaction is chosen to be $V=2$, although we
have explored other values.  The hopping matrix elements to both
nearest and second-neighbor sites are chosen to be $t=t'=1$,
although again we have explored other values.  The second-neighbor
hopping is included so that the model remains nonintegrable (quantum
chaotic) and thus diffusive at zero randomness \cite{moh}. The
on-site potentials $w_i$ are independent Gaussian random numbers
with mean zero and variance $W^2$. Each realization of the disorder
potential will generally have mean-square random potential
$\sum_{i=1}^L w_i^2/L$ that is not precisely $W^2$. We have found
that restricting our ensemble of samples to those with mean-square
random potential {\it precisely} $W^2$ reduces our statistical
uncertainties by about a factor of 2 in the largest samples. This
change of statistical ensembles does produce quantitative changes in
the spectral properties (mostly noticeable for intermediate values
of disorder, $4\lesssim W \lesssim 9$), but it does not appear to
produce any qualitative changes in the finite-size scaling behavior
that would affect our conclusions and it cannot affect the system's
intensive properties in the thermodynamic limit.

We study all many-body eigenenergies of this Hamiltonian, weighting
them equally; the data shown here are for half-filling, $L/2$
particles. Thus we are studying temperatures high compared to the
energy scales of this Hamiltonian. If a localized phase does indeed
exist in this model, it should be present even at high $T$ for
strong enough disorder. An important motivation for this choice of a
model was our recent work \cite{moh} on the same model in the
absence of randomness, where the approach to thermodynamic limit was
rapid enough to observe the onset of hydrodynamic behavior with
$\leq 9$ particles.  Here we are limited to somewhat smaller sizes,
since the random potential violates momentum conservation; we focus
on sizes up to $L=16$. The number of realizations needed to achieve
adequate statistical certainty depends strongly on $W$ and even more
so on $L$. At $L=8$ we average over $10,000$ realizations whereas at
$L=14,16$ only about 50 to 100 suffice \emph{except} in the putative
critical region, where we average over $1000$ realizations for each
$W$.

\section{Method of analysis}

To look for the diffusive-to-insulating phase transition in this
model, we have chosen to use what appears to be numerically the most
accessible quantity that shows a clear, well-understood difference
between the two phases, namely the spectral statistics of adjacent
energy levels of the many-body Hamiltonian.  In the localized,
insulating phase (assuming it exists in our many-body system), in
the thermodynamic limit of a large sample, the eigenstates are
localized in the many-body Fock basis of localized single-particle
orbitals, so states that are nearby in energy are far apart in this
Fock space and do not interact or show level repulsion.  As a
result, nearby energy levels are simply Poisson distributed
\cite{PoissonCaveat}. In the diffusive phase, on the other hand, the
level statistics of a large sample are those of random matrix
theory, the Gaussian orthogonal ensemble (GOE) in particular. For
the finite-length samples that we can diagonalize, the level
statistics cross over smoothly between these two limiting behaviors
as the strength of the random potential is varied. This crossover
becomes sharper as the length $L$ is increased, and we can look for
a phase transition using standard finite-size scaling techniques;
this approach works well for the single-particle localization
transition in three dimensions (see, e.g., \cite{sssls}).

The choice of a quantity to compute and use for the finite-size
scaling analysis is to some degree arbitrary: the hypothesis of
universality implies that many features of the distribution of
eigenvalues of the Hamiltonian are universal in the thermodynamic
limit \cite{PoissonCaveat}.  By analogy to the Binder ratio for
phase transitions with a local order parameter \cite{binder}, we
seek a dimensionless measure of spectral statistical properties, say
$r(W,L)$, that is expected to take different finite values in the
thermodynamic limit, $L\rightarrow \infty$, in the two phases and at
the critical point ($W>W_c$, $W<W_c$ and  $W=W_c$).  Since the zero
of energy is arbitrary, it is natural to work with gaps between
many-body levels. Here in particular we consider gaps between {\it
adjacent} many-body levels,
$$\delta_n=E_{n+1}-E_n\geq 0~,$$ where the eigenvalues of a given
realization of the Hamiltonian for a given total number of
particles, $\{E_n\}$, are listed in ascending order.

The dimensionless quantity we \cite{alternatives} have chosen to
characterize the correlations between adjacent gaps in the spectrum
is the ratio of two consecutive gaps
$$0\leq
r_n=\min\{\delta_n,\delta_{n-1}\}/\max\{\delta_n,\delta_{n-1}\}\leq
1~.$$ For uncorrelated Poisson spectrum the probability distribution
of this ratio $r$ is $P_P(r)=2/(1+r)^2$, and its mean value is
$\langle r\rangle_P=2 \ln 2 -1 \cong 0.386$.  The
numerically-determined probability distribution \cite{goedata} for
large GOE random matrices is shown in Fig. 1; its mean value is
$\langle r\rangle_{GOE}=0.5295\pm 0.0006$. As expected, level
repulsion/spectral rigidity in the GOE spectra manifests itself in
the vanishing of the probability distribution $P_{GOE}(r)\sim r$ as
$r\rightarrow 0$.
\begin{figure}[h!]
\begin{center}
$\begin{array}{c}
\includegraphics[width=5.3cm,angle=-90]{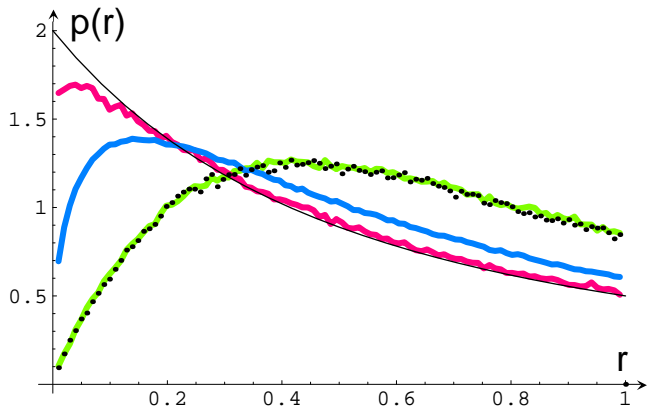}
\end{array}$
\end{center}
\caption{(color online)  Disorder averaged probability distribution,
$P(r)$, for Poisson (solid black line) and GOE distributed
eigenvalues \cite{goedata} (black dots) and for our interacting
fermion model at randomness $W=3$ (green, diffusive regime), $W=11$
(red, localized regime) and $W=7$ (blue, intermediate) for length
$L=16$.} \label{fig:pofr}
\end{figure}

\section{Results}
With interaction and hopping terms fixed as above $(t=t'=V/2=1)$, we
vary the strength of disorder from $W$=1 to $W$=10 or more and for
each $(L,W)$ we diagonalize a large number realizations, $R$ (see
above). For each sample we compute the spectral average of $r$,
$\langle r \rangle$, over \emph{all} states. We then
disorder-average this quantity, $[\langle r \rangle ]$, to arrive at
$r(W,L)$ exhibited in Figure 2. The statistical uncertainties in
$[\langle r \rangle ]$ are estimated as usual as $\pm (([\langle
r\rangle^2]-[\langle r\rangle]^2)/(R-1))^{1/2}$.
\begin{figure}[h!]
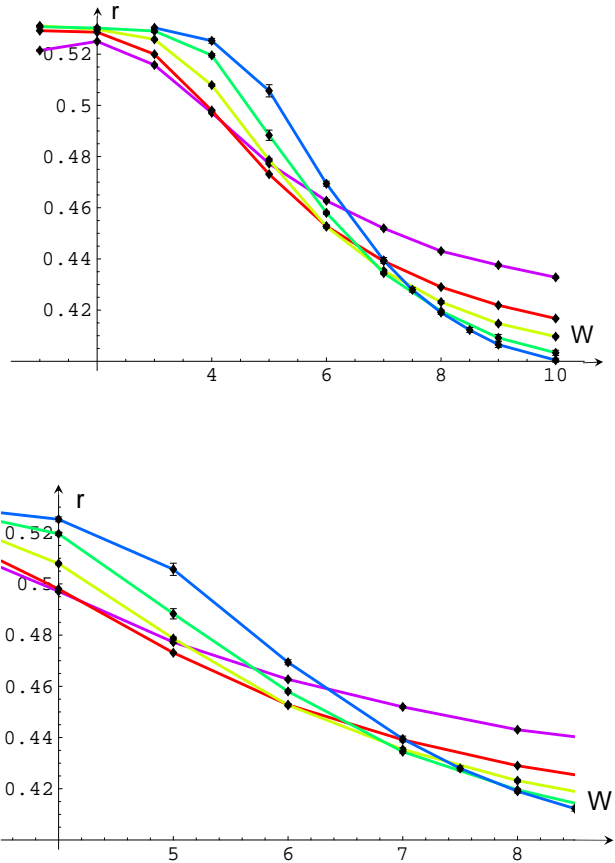

\begin{center}
$\begin{array}{c} \epsfxsize=3.1in
\includegraphics[width=5cm, angle=270]{Fig2b.epsi}
\\
\\
\\
\\
\includegraphics[width=5cm, angle=270]{Fig2a.epsi}
\end{array}$
\end{center}
\caption{(color online)  Size $L$ and disorder $W$ dependence of
$r(W,L) $. The curves correspond to $L$=8, 10, 12, 14, 16 from top
to bottom for large $W$. Bottom: an enlargement of the crossing
region to make the drift of the crossings more visible.  Where not
visible, the error bars are smaller than the points.}
\label{fig:pofr}
\end{figure}

As expected, larger samples have more Poisson-like statistics than
smaller ones for strong disorder, $W>8$, in an apparently localized
regime; while for weak disorder, $W<4$, the level statistics
converge towards GOE with increasing $L$, since this is the
diffusive phase. We have checked that the entire probability
distributions $P(r)$ in these regimes approach those of Poisson and
GOE spectra (see Fig.1). There is an additional crossover at very
weak disorder: as crystal momentum conservation is recovered there
appears a turnaround in the statistics as the decoupling of
different momentum sectors suppresses the average of $r$ below its
GOE value. This latter crossover at weak randomness is a nuisance
for us and we steer clear of it the best we can by working away from
the clean limit and also not considering very short chains (with
less than 8 sites) where this momentum pseudoconservation persists
to larger values of disorder: $r(W,8)$ shows a remnant of this
crossover at $W=1$, while larger values of $L$ do not show it at all
over the range of $W$ considered.

The simplest one-parameter finite-size scaling scenario for the
proposed diffusive-to-insulating phase transition would have these
traces of $r(W,L)$ vs. $W$ at fixed $L$ in Fig. 2 all cross at $W_c$
as $L\rightarrow \infty$, with a slope that increases with
increasing $L$ (e.g., see \cite{sssls}). However, we find that the
crossings of the $r(W)$ curves for adjacent $L$'s take place at
points that, as $L$ is increased, ``drift'' progressively towards
larger $W$ and smaller (more insulating) $r$; see Fig. 2. As this
drift precludes the straightforward quantitative analysis of our
data in terms of one-parameter scaling theory, we have exerted
considerable effort to attempt to eliminate it \cite{IrrelevantOps},
including looking at other temperatures, interactions and fillings,
candidate scaling variables other than $r$, and selective spectral
averaging (e.g. excluding states in the high and low energy tails of
the spectrum). While this drift of the crossings can be reduced
(particularly by trimming tails or reducing the temperature) it
appears that it is intrinsic to this model's spectral statistics and
none of the many things we have tried eliminated or reversed it.
Accepting this, there are two very distinct possible implications:
either the drift converges to a finite $W_c$ (and likely to
$r_c=\langle r\rangle_P$; see below) in the large $L$ limit, or it
continues indefinitely to $W_c=\infty$ which would imply that the
insulating phase does not exist at these high temperatures. In fact,
this latter possibility has already been advocated in previous work,
see e.g. ref. \cite{ss00}, where it was argued that $W_c(L)\sim L$
(i.e. $W_c(L)$ is a length dependent scale at which spectral
statistics changes from Poisson-like to GOE).

Although at this point we cannot choose between these two
possibilities based on these data for the spectral statistics, it is
worth making some more comments about the former possibility
\cite{IrrelevantOps}: The apparent drift of the crossing points,
$\{W_c(L), r_c(L)\}$, is indeed substantial along the
\emph{vertical} axis, as would be expected if $r_c$ is converging to
the Poisson-limit value $\langle r\rangle_P$. Thus these data seem
consistent with a large $L$ limiting behavior $W_c(L)\rightarrow
W_c<\infty$, $r_c(L)\rightarrow \langle r \rangle_P$, whereby the
critical point is insulating as far as level statistics are
concerned. There are independent reasons, based on analogy to
Anderson localization on high-dimensional and Cayley graphs
\cite{CayleyHighD}, to expect such a behavior. The proposed
many-body localization transition is a localization transition in an
infinite-dimensional Fock space \cite{baa}. Given that, there may be
plenty of room in that space for the states at a
diffusive-to-localized transition (i.e., at the mobility edge) to
have an infinite localization length but still have a negligible
overlap between states and thus no level repulsion and Poisson level
statistics.  This would imply that the spectral statistics should
converge to Poisson as $L$ increases both within the localized phase
and at the transition, and thus the ``crossings'' in our Fig. 2 must
move down to $r=\langle r\rangle_P$ in the large $L$ limit.  This
scenario, with a localized phase for $W>W_c$, seems qualitatively
consistent with the data we have presented above. Unfortunately, if
this is indeed the case then spectral statistics are not a good tool
for simple finite-size scaling analysis.  We shall explore other
approaches to this problem in the near future.

\section{\label{sec:conclusions}Summary and outlook}
We have looked for signatures of the proposed many-body localization
transition in the statistics of exact spectra of a one-dimensional
tight-binding model of strongly-interacting spinless fermions in a
random potential. Although some indications of this phase transition
are clearly seen, there are rather strong deviations from and/or
corrections to finite-size scaling present. The latter might be
interpreted as calling in to question the existence of the proposed
many-body localized phase at the high temperatures we study.
Alternatively, this failure of simple one-parameter finite-size
scaling might be because the critical point has insulator-like
spectral statistics.

In closing, it may be worth noting that thus far we have focussed on
the most elementary aspects of many-body localization. These may not
be necessarily the easiest to study experimentally. Finite-size
effects in dynamical response functions, i.e. conductivity, appear
more delicate but they are certainly worthwhile understanding, as
data may already exist in regimes of interest, in materials as
diverse as magnetic salts and disordered conductors.

\begin{acknowledgments}
We are grateful to D. Basko, I. Aleiner, B. Altshuler, A. Lamacraft,
S. Kivelson, A. Garcia-Garcia, N. Read, S. Girvin, Y. Alhassid, S.
Sondhi, S. Mukerjee and S. Sachdev for discussions. The authors
would also like to thank the NSF for support through DMR-0603369
(V.O.) and MRSEC grant DMR-0213706 (D.A.H.).  V. O. is also
supported by a Yale Postdoctoral Prize Fellowship.
\end{acknowledgments}


\begin{thebibliography}{99}
\bibitem{goedata}
GOE data was obtained from $1000$ random matrices of size $3432$,
which is the number of states in a half-filled chain of $L=14$.
\bibitem{IrrelevantOps}
The simplest one-parameter finite-size scaling would have the
spectral statistics at the mobility edge intermediate between
Poisson and GOE, as appears to be the case for single-particle
localization (e.g., Ref. 14).  In this case, drifting crossing
points would be due to slow decay with increasing $L$ of effects
from operators that are irrelevant (in the RG sense) at the critical
fixed point.  It should be possible then to eliminate or even change
the sign of these effects by either changing the observable being
studied or the Hamiltonian itself, to cross the phase boundary where
the effects due to the irrelevant operators are either
quantitatively small or of the opposite sign.  However, it is
possible that the sign of the irrelevant perturbation is constrained
by physical considerations, e.g. in scalar $\phi^4$ theory the
quartic term is irrelevant above four dimensions but must be
positive for the low temperature phase to exist (the vortex fugacity
at the Kosterlitz-Thouless transition in another example, but there
it is the high-$T$ phase that needs it). We discuss (in main text)
how something along these lines, that would invalidate the simplest
one-parameter scaling analysis, might be responsible for the drift
of our crossings in Fig. 2, and for the reason that we have been
unable to eliminate or reverse this drift by varying the model or
the quantities we measure.
\end{thebibliography}

\end{document}